# Mobile Information Collectors' Trajectory Data Warehouse Design

Wided Oueslati[1] and Jalel Akaichi[2]

[1]Department of Computer Science, High Institute of Management, Bouchoucha, Tunisia
`widedoueslati@live.fr`
[2]Department of Computer Science, High Institute of Management, Bouchoucha, Tunisia
`jalel.akaichi@isg.rnu.tn`

## ABSTRACT

*To analyze complex phenomena which involve moving objects, Trajectory Data Warehouse (TDW) seems to be an answer for many recent decision problems related to various professions (physicians, commercial representatives, transporters, ecologists …) concerned with mobility. This work aims to make trajectories as a first class concept in the trajectory data conceptual model and to design a TDW, in which data resulting from mobile information collectors' trajectory are gathered. These data will be analyzed, according to trajectory characteristics, for decision making purposes, such as new products commercialization, new commerce implementation, etc.*

## 1. INTRODUCTION

In order to make the best decisions in the right time, and to act efficiently, the decision maker must have on the one hand a set of sufficient, available, reliable, relevant, precise and recent data and on the other hand a set of subject-oriented, integrated, time-variant and non volatile data. That leads us to say that information collection and analysis is essential for decision makers, working for or owning enterprises, willing to invest in new projects related to various and complex domains. In fact, investors need to gather for analysis, information about the environment where the investment may take place to measure the chances of success or failure. For example, an investor in the tourism domain must have an idea about the site where he may build or rent a hotel, the infrastructure of roads leading to the site, the communication infrastructure, the air pollution surrounding the place, the climate conditions of the zone, the availability of labour in the tourism domain in the town, the existence of academic centres related to tourism, etc . This can be ensured and enhanced, in one hand by recent and incredible evolution of positioning technology, mobile devices, wireless networks, etc, and on the other hand by business intelligence techniques such as data warehouses and data mining.

In fact, these technologies make the collection of data and their updates feasible and in real time while moving. Moreover, these new technologies contribute to the creation of new jobs related to the collection, the management and the storage of data in order to have them exploited for analysis in depth, by new customers such as investors and of new opportunities seekers.

The storage and integration of huge amount of heterogeneous data in order to be exploited at the analyze phase, must be in an appropriate repository which facilitates decision support system applications. In our research work, Mobile Professionals (MPs) are in charge of the information collection. They are equipped of mobile devices, and transportation means equipped by sensors. They move through multiple trajectories and generate a large amount of data ensuing from planned and non planned observations. Such data ensuing from trajectories of moving objects are so called Trajectory Data (TrD). Those latter are huge, variable in space and in time and heterogeneous as they come from different sources. They need to go through a phase of cleaning and integration in order to be gathered into a repository suitable for modelling the continuity of changes at the level of space and time, for querying and for analysis. In addition, the data organization is not always adapted to the aimed subject of analysis and need an appropriate modelling and transformation. As we know the STDW techniques aim to the integration of spatial data which comes from heterogeneous sources and their transformation into a model that facilitates the application of analysis tools. The limitation of the STDW is the discreet representation of the two dimensions space and time, while the trajectory data of a moving object is known by the continuity in space and time levels. This leads to consider that the Trajectory Data Warehouse (TDW) as the best mean which permits the modelling and the analysis of trajectory data in a multidimensional context. In fact, TDW allow decision makers of different domains to model and to query their moving objects complex data by using specific concepts and tools. It is used as a common framework for validation, reproduction and simulation of experimental data analysis.

The goal of this paper is to make trajectories as a first class concept in the trajectory data conceptual model and to propose a conceptual model for the TDW destined to gather trajectory data into a repository for query analysis triggered by On Line Analytical Processing (OLAP) users. The challenge is to design a conceptual model for trajectory data, trajectory data warehouse and languages that allow us to formulate queries in a simple and precise way.

This paper is organized as follows: In section 2, we will present different research works related to movement scenarios of moving objects, trajectory data conceptual model and the data warehouse modeling (Data Warehouse, Spatial Data Warehouse, Spatio-Temporal Data Warehouse and Trajectory Data Warehouse). In section 3, we will propose a trajectory data conceptual modeling for mobile professionals and especially for Mobile Information Collectors. In section 4, we will propose a trajectory data warehouse conceptual modeling for our illustration case with the snowflake schema. In section 5, we will summarize the work and propose new perspectives to be done in the future.

## 2. RELATED WORKS

In this section we will focus in the literature on the study of the movement of moving objects, the trajectory data conceptual modeling, the trajectory data warehouse conceptual modeling and queries on moving objects.

### 2.1. Moving objects' movement scenarios

With the development of localization systems and wireless networks, new applications based on mobile sources generate a huge amount of data. Mobile sources often called mobile objects are spatial objects of which the shape and/or the localization change continuously during the time. In fact, many applications dealt with the study of movement of mobile objects such as traffic evolution [1], migration of birds [2], etc. Some others are worth being to be investigated such as natural disasters related applications (forest fires, floods, hurricanes, etc). In the literature, most of researches work focused on moving objects in the field of databases [3-4] and few works were interested in the data warehouse field [5-6] where huge data can be stored and analyzed for decision making goals. Whatever the field taken into account, researchers distinguish three movement of objects scenarios [5]:

- Unconstrained movement or free movement (such as floods, hurricanes or forest fire, etc. in fact, such objects move arbitrarily without any constraint. It is a natural phenomenon).
- Network-constrained movement (such as cars, plane and train. In fact, the train can move only in a specific network which is the rail).
- Constrained movement (such as pedestrians, birds, etc. In fact, there are some kinds of birds that migrate at a given season. The movement of the birds is constrained by the season of migration and sometimes by the weather. For the pedestrian, he can not walk for example in a private zone).

For the unconstrained movement of a moving object, we found in the literature the model proposed by Meng and Ding [3]. They propose an approach named DSTTMOD (Discrete Spatio-Temporal Trajectory Moving Object Database system) destined to model the free movement of moving objects.

The researchers supposed that the moving object moves in a space of two measurements(X, Y) and with a constant speed, nevertheless this condition is not realistic. The trajectory is represented by a curve (X, Y, T) where the couple (X,Y) represents the space dimension and (T) represents the time dimension. The DSTTMOD model supports some user's query to determinate the past, current and future localization of a moving objects. In fact, the future localization is given in advance. Nevertheless, it is very hard to prevent some thing like this in real world in some cases such as forest fires, floods, hurricanes, movement of pollution…

For the modeling of constrained movement by the network, we will present the works of Wolfson, Ding, Pfsoer and Guting.

Wolfson and Chamberlain were the first to propose a model for the constrained movement of moving objects by the network in the setting of project so called Databases fOr MovINg Objects tracking (DOMINO). In fact, they defined a model for data and for predicates in order to enhance query on objects which are moving in a network. In this kind of mobility, researchers model not only the moving object but also the network. In fact, researchers in [8] describe the network by a relational model where every tuple of the relation represents a section of road between two intersections (junctions). The network is characterized by a set of attributes.

In [8], the trajectory is seen as a sequence of sections of roads crossed continuously by the moving object. The details of each trajectory are obtained while using the shorter path algorithm and giving the input data at the entry of each trajectory. Those inputs data are: the departure address, time of departure and the arrival address. The speed is constant along each section of the trajectory.

Ding and Guting [9] were interested in the managing of objects on the dynamic transportation network. In this model called MODTN, the moving object is considered as a point moving on a predefined network graph. The transport network is modeled by a dynamic graph. This latter takes into account the traffic jam, construction projects as, the transport network has a changed state (traffic jam, breakdown) and topology (construction project, insertion or delete of junction). The moving object changes of localization and/or shape, continuously in time. Mobile object and transport network are both dynamic. To model the dynamic graph, researchers propose to associate a temporal attribute for each road or junction in order to detect and store their state for each moment. Roads are represented by polylines with arbitrary geometric shape.

In [6] [7], Pfoser and Jensen proposed to transform the network and the trajectories in order to reduce their dimensionality. In fact, they transform the 2D representation to a 1D representation for the network: the segment of the network are ordered according to a succession of sections, then the obtained sections are transformed to intervals and ordered one after the other in an axis of one dimension.

After modeling the network in 1D, researchers modeled trajectories in the new network. This step consists on projecting segments and their lengths in the network, this succeeds to a whole of intervals, and trajectories are presented in a 2D space while adding their temporal extensions.

In [4], Guting and Al consider that the network is composed by a set of roads and junctions.A road is defined by a set of attributes such as: identifier, geometry, length and type.

The attribute "type" can have two values: "simple" (if the road has a unique sense) and "dual" (if the road has a double sense).

A junction is considered as an intersection between two roads. Each junction is defined by his position relative to each road and by a matrix which represents the possible circulations.

To model trajectories, researchers define applicable operations at fixed instances, and then define a process which allows transposing those operations at mobile instants. For example, it is possible to define a dynamic point moving continuously on the network from a static point while applying the mpoint operator (MovingPoint).

For the constrained movement, the object moves with some constraints other than the network. We found in the literature the work of spaccapietra [2]. This latter, was interested in the study of the migration of birds. The movement of such object is constrained as we know by the weather, the season and perhaps other conditions.

In [2], the moving object is the white stork. This latter is characterized by some attributes such as the name, the birth, the year, the location, the death, the year, the location…

The trajectory of the white stork is of type spatio-temporal and it is represented by a position of begin, a position of end and a set of stops and moves between those positions. The stops are defined as stable points in a given time interval, while the moves are defined as variable points in a given time interval. Both stops and moves have a set of attributes.

### *2.2. Trajectory data conceptual modeling*

Many researchers were interested in the conceptual modeling of data with the E\R diagram [2] [12-13] or with UML diagram [10-11], but few of them were interested in the conceptual modeling of trajectory data, perhaps because the notion of trajectory is a new concept. Trajectory data includes data specific to moving objects such as professionals on the move, migration birds, natural phenomena, etc. A big interest was shown in the managing of moving objects [8] [14-15], but a very little interest was revealed in the concept of trajectories of moving objects at the conceptual level. At the best of our knowledge, the first work that was interested in trajectory is presented in [16]. It tried to enrich the moving object model with some semantic annotations. These latter include properties of the moving object such as movement speed, direction, covered area… It also introduces a relationship between trajectory and its spatial environment (stay within, by pass, leave, enter, cross), and a relationship between trajectory and other spatial objects (intersect, meet, equal, near, far).

The focus was then changed from the formalization of the mobile objects to the management of their trajectories in order to derive behavioral patterns through moving objects trajectory data analysis. These patterns can be helpful to prevent some good measures in a specific domain and to simulate models [2][17-18]. In fact, analyzing the trajectory of moving objects, will allow the researchers to induce measures for decision making.

The two approaches of conceptual modeling proposed in [2] (trajectory design pattern and trajectory data type), were driven by different modeling goals and can be combined if needed.

The design pattern approach: the design pattern is a predefined generic schema that can be connected to any other database schema by the designer and can be modifiable. In fact the designer can modify the design pattern by adding new elements or semantic attributes, deleing other ones to adapt it to the requirements of the new application [2].

The Data Type approach: this approach hold the idea that many semantics information are specific to application and can not be collected into a data type, but has to be defined by the database designer.

On the other hand, the authors propose to define a generic data types to hold the trajectories' components such as the begin, the end, stops and moves… and to define the functions of interpolation.

Among the data types defined in [2], we cite:
- Data type TrajectoryType: each element of TrajectoryType describes a single trajectory and it is composed of a time varying point to represent the spatio-temporal trajectory, a set of sample points, a list of stops, a list of moves.

- Data type TrajectoryListType: each element of TrajectoryListType describes a list of trajectories which is composed of a list of element of type TrajectoryType.

Each Data Type supports a set of methods. In fact, for the TrajectoryType we found several methods like begin( ), end( ), SamplePointsVarying( ), StopAt(t Instant), AverageSpeed( ),

NumberOfStops( ), Orientation(n Integer)… and for the TrajectoryListType we found two methods which are TrajectoryAtInstant(t Instant) and DurationBetween(n Integer).

After presenting different research works in the field of moving objects, we propose to compare between different models of moving objects resulting from studies cited above. To do this we propose the following comparative table.

| | **Type of movement** | **Modeling of the space of movement** | **Modeling of the trajectory** | **Network constraint taken into account** | **Dynamism of network taken into account** | **Variation of speed taken into account** |
|---|---|---|---|---|---|---|
| **Model of Ding and Meng** | Free movement | Modeled by a set of curved in the space | Modeled by absoluted coordinates | No | No | No |
| **Model of Guting** | Network constrained movement | Modeled by graphs on the transport network | Modeled by relative positions on the road | Yes | Yes | Yes |
| **Model of Wolfson** | Network constrained movement | Modeld by a set of sections | Modeled by absoluted coordinates | No | No | No |
| **Model of Pofser and Jensen** | Network constrained movement | Modeled by transformed segments in 1D space | Network constrained movement | No | No | Not treated |
| **Model of Guting and Al** | Network constrained movement | Modeled by roads and junctions | Modeled by relative positions on sections of the road | No | No | Not treated |
| **Model of Spaccapietra** | constrained movement | Modeled by geometric forms | Modeled by a set of stops and moves | No | No | Not treated |

Table 1. A comparative study between models related to moving objects

### 2.3. Trajectory data warehouse conceptual modeling

In the last years, there have been several approaches to design Data Warehouses from the logical, physical and conceptual perspectives. The conceptual model represents the human understanding of a system. In fact, a conceptual model gives to user information about the modeled reality. That information will help the user to understand data and make ad-hoc queries.

As mentioned in [26], traditional conceptual models for database modeling such as the Entity-Relationship model [27], the extended Entity-Relationship model with the GMD (Generalized Multidimensional Databases) [28] (in this model, authors combine the well known first order semantics of the standard E\R [29] with the model theoretic semantics of GMD [30-31]) and the StarER model [27] (in this model, authors combine the structure-efficient star schema with the semantically rich E\R model) are not suitable to describe multidimensional aspects of a DW. For those reasons several multidimensional models have been emerged in the context of databases and especially DW [32-33].

Multidimensional modeling is characterized by two primitives which are Facts and Dimensions. Those latter are used to construct the star schema [34-35], the snowflake schema [36-37] or the constellation schema [38-39]. Many applications need to integrate spatial and temporal data types. For this purpose a Spatio-Temporal Data Warehouse (STDW) [19-20] technology may be very helpful, but in fact current STDW is not sufficient to deal with complex objects such as continuous field, object with topology, moving object in different domains such as telecommunication [21] ( cell-phones trajectories), ecology [22] ( analysis of animal's behavior through its trajectories), biology( proteins migration in human body [23], Naïve-T cell [24], CD4 cell [25] ). To solve this problem a Trajectory Data Warehouse (TrDW)[21-22-23-24-25] was setting up.

## *2.4. Queries on moving objects*

A lot of investigations were done in the field of querying moving objects in databases and in data warehouses. In fact, in [40], Wolfson and Chamberlain have proposed a model to know the current and the near future position of moving objects. In [41], authors extend the GeoPQL (Geographical Pictorial Query Language) [42] by introducing temporal operators such as: T-before, T-meets, T-overlap, T-stats, T-during, T-finishes, T-equal and Spatio-temporal operators such as Geo-growing, Geo-shrinking, Geo-translation, Geo-merging, Geo-splitting, Geo-appearing, to represent spatio-temporal queries In [43], authors present new types of spatio-temporal queries and algorithms to process those. They considered that the moving objects are point objects. They defined two types of queries. In fact, we found the coordinate-based queries which have as operations: overlap, inside, ect and trajectory-based queries, those latter can be classified into two sub types which are: topological queries and navigational queries.

Enriching OLAP operators and specifying OLAP language for querying and manipulating trajectory data was done in [44]. So we found the basic OLAP operators such as: Operators for clustering of trajectories, Operators for modifying the spatio_temporal granularity of measures who represents trajectories, Operators to construct automatically trajectory data marts, Operators for aggregating imprecision present in the data of the TrDW, Operators for incremental data analysis on trajectory data (this kind of operators is essential to analyze the mobility).

## 3. TRAJECTORY DATA CONCEPTUAL MODELING FOR MICs

The development of new technologies such as: sensors, GPS, PDA played an important role in the emergency of new professionals. It is called Mobile Professionals. Those latter can be mobile physicians, mobile commercial representatives, mobile collectors of information.

In our model, the moving object is the Mobile Information Collector (MIC). His mobility will be constrained by the network. In fact, we suppose that the company, involved in collecting information about various commercial and investment domains, useful for decision making, about diverse domains. To perform an assigned task, this company has three teams: a team formed by the MICs to collect data, a team formed by the decision makers to make decision after analyzing the trajectory data and teams formed by the responsible of missions to select and appoint the first team. In fact, the second team hires many MICs and plans for every one his trajectory and some tasks to do and give him a means of transportation (e. g. a car), eventually, equipped with some sensors (e. g. pollution sensors). Moreover he is outfitted with a PDA, connected with our mediator, used for trajectory data inputs and storage. The use of the PDA is very helpful for the mission of the MIC, since it provides in standard a note book of address, a software of messaging making possible the consultation of mails as well as the writing of new messages thanks to the GPRS (General Packet Radio Service) network, and a GPS receiver permitting to take advantage of multiple services of geo-localizations. While moving in some delegations or some regional governments the MIC may meet various natural or artificial point of interest. The MIC may stop for personal, planned, or unforeseen reasons. The move of the MIC must be in some sections of the trajectory.

## 3.1. Mobile professionals' modeling

With the development of technology of communication, many professionals become mobile in different domains such as commerce, medicine, education… For this reason, we propose a generic conceptual model for mobile profesionals'trajectory data at a first time then adopted it to our case study which is the mobile information collector.

Each mobile professional is equipped by a PDA and has a trajectory which is composed of a set of trajectory sections. Those latter are formed by a set of moves and stops. A given move is delimited by two stops. This latter is around points of interest.

Let's mention that points of interest defer from one mobile professional to another (for example for mobile physicians points of interest are hospitals, emergencies, patients' houses…) .

The following figure represents the generic conceptual model for trajectory data resulting from mobile professionals.

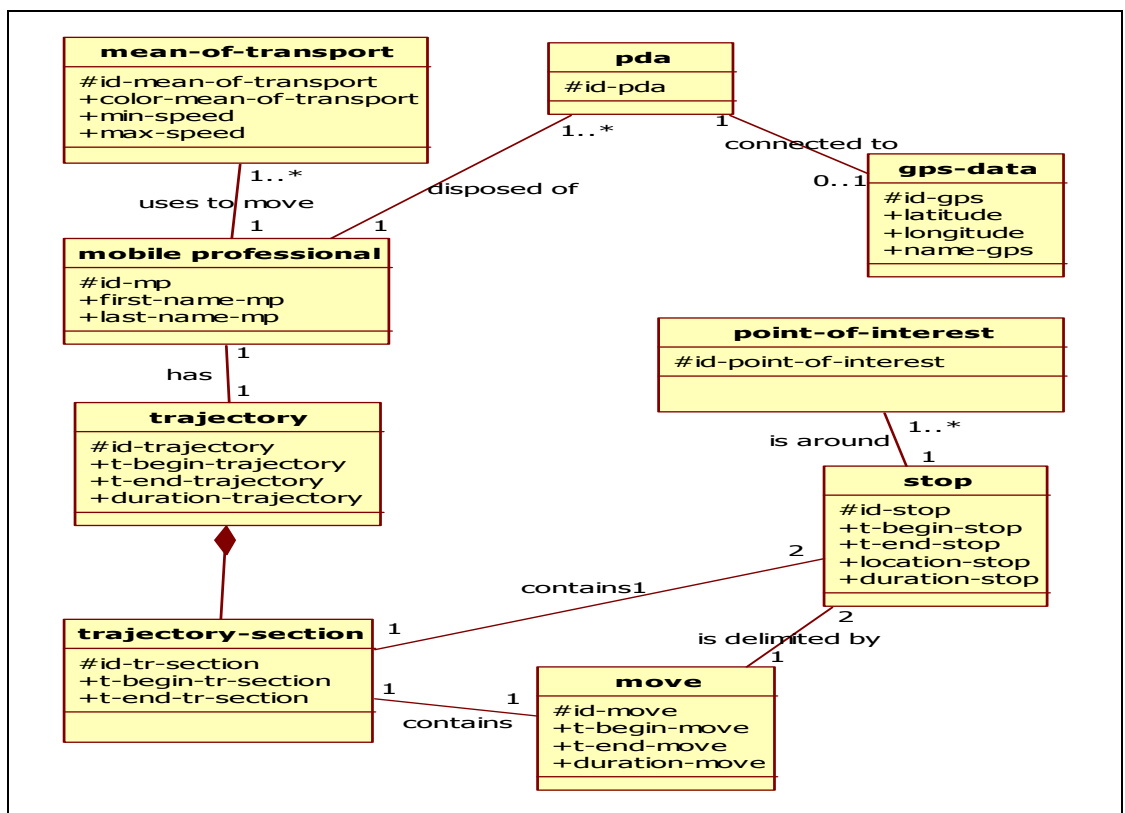

Figure 1.  Generic conceptual trajectory data model

## 3.2. Mobile information collectors' modeling with STAR UML

we have to identify at first, the MIC's trajectory data conceptual modeling,  at a second step, the MIC's itinerary diagram and at a third step, the MIC's trajectory description.

### 3.2.1. Conceptual trajectory data modeling

Mobile Information Collectors move in search of better opportunities for implementing new commerce installations, finding new markets for existing and new products … For our application, MICs were equipped with PDAs on which we installed a Graphical User Interface (GUI) responsible of collecting and transmitting data. This GUI is related to a mediator in charge of ETL (Extraction, Transformation, and Loading) operations. Hence ETL operations performed, analysis and mining tasks can be performed in real time.

Analyzing MIC's motion helps to get better knowledge not only about MIC's work and behavior but also about their environment in which they move: roads infrastructure, weather conditions, transport means states, hazards encountered …

The conceptual model for trajectory data resulting from the mobile information collectors is presented by an UML Class Diagram (see figure8). We will deal with the modeling issue at the conceptual level, as this is the level best suited to represent application requirements. The conceptual level is insensitive to technological changes and drives logical design towards a correct solution.

For our conceptual representation we will use the class diagram of star UML because of its popularity and expressiveness.

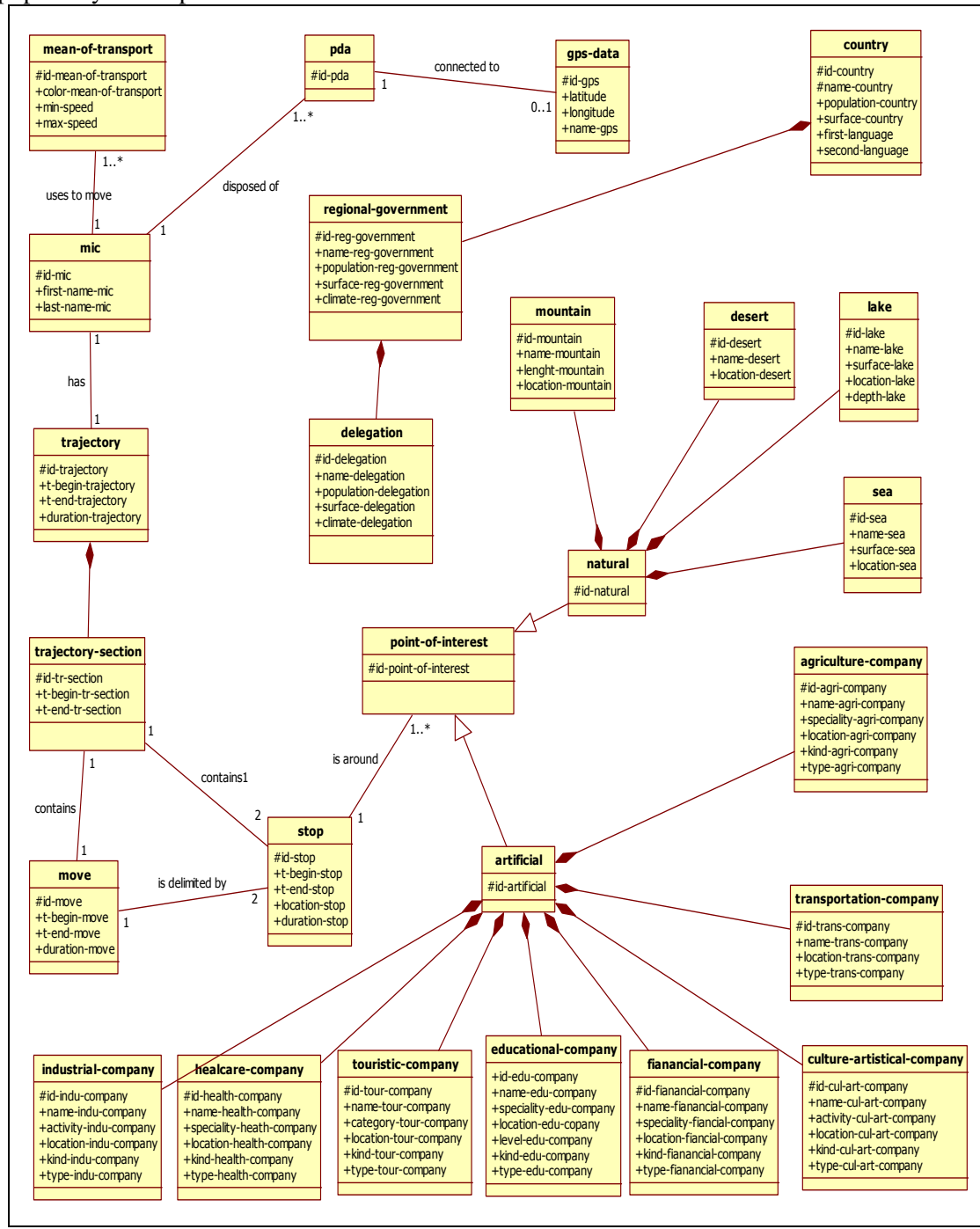

Figure 2. Class diagram for MICs' trajectory data

We will describe the main classes of the MIC's trajectory data and the relationships between them.

- **Mobile Information Collectors**

We suppose that the company, involved in collecting information about various commercial and investment domains, hires many Mobile Information Collectors represented by the MIC Class at the conceptual level. A MIC is identified by an identifier and has some attributes such as first-name-mic, last-name-mic.

In order to understand the properties of a MIC, we propose to model his internal structure and his characteristics. In fact, in the real world, the MIC uses PDA to send data. He has to identify with an authentication key. He has three attempts for authentication. This forms the history of authentication for a given MIC. This latter will be valued by the head of the mission. Such valuation concerns the moving capability, the communication capability and general knowledge. Each MIC has some tasks to do at some places and can meet some navigation events. This forms the itinerary of the MIC.

| Mobile Information Collector Class name |
|---|
| "Attribute": id-mic, first- name-mic, last-name-mic |
| " Authentication": mic's authentication key |
| "History": success , error |
| "Capability": moving capability, communication capability, knowledge |
| "Itinerary": set of destinations, set of tasks, navigation events |

Table 2. The MIC's internal structure

- **Move**

Each MIC moves through a given set of sections of his trajectory using a given mean of transport such as a car. In his trajectory, he has to annotate points of interest while stopping.

Each move has a geometry which is a time varying point and a life cycle which is a simple time interval. This latter is represented by the "duration" attribute which is obtained while applying the following formula:

Duration move =life cycle move= (time.end.move – time.begin.move)

Each move is delimited by two successive stops.

- **Stop**

A stop is a semantically important part of a trajectory where is considered that the object has not effectively moved [36]. Each Stop object has a life cycle which is a simple time interval and a geometry which is a "point" and a kind, in fact in our definition we considered that we have three types of stop: a planned stop in order to observe and to collect and then to send information, a private stop in order to have a pause and to eat or something like this and finally an unforeseen stop when there is some navigation events like a breakdown or a bad weather.

A given stop is in a given section of the trajectory. Each section of the trajectory is represented at the conceptual level by the so called trajectory-section class.

- **Trajectory-section**

It's a part of a trajectory. Each trajectory-section is composed by two stops and one move.

The trajectory- section class has an identifier and a set of attributes like t-begin-section, t-end-section.

- **Trajectory**

Each MIC has a trajectory fixed by the responsible of the mission. Semantically a trajectory is defined as an ordered set of stops and moves [20], since it is the sum of a set of trajectory-section.

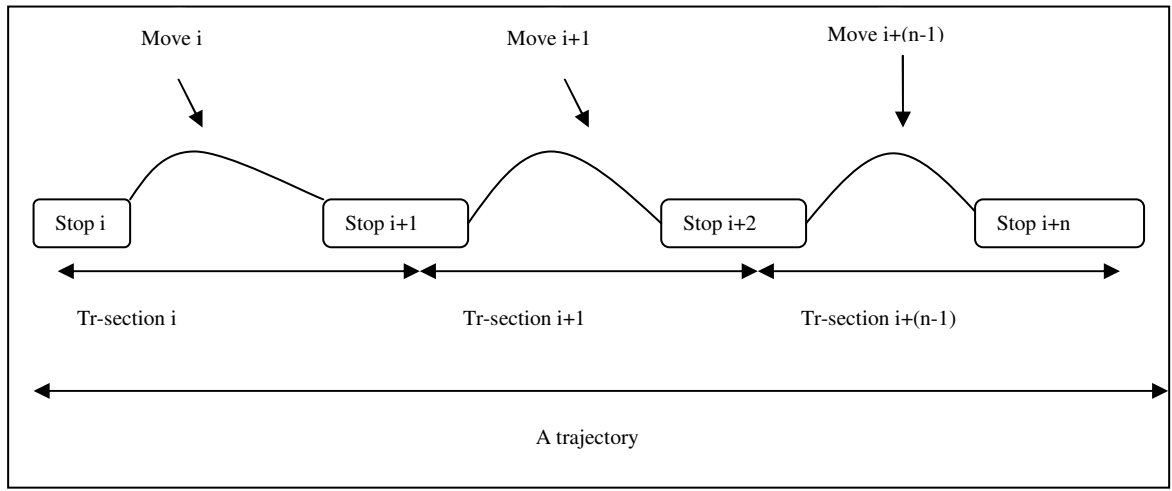

Figure 3. A trajectory

- **PDA**

Each MIC has one PDA to send trajectory data and to communicate with the responsible of his mission. The PDA has an identifier and it is connected to a GPS.
- **GPS-data**

The GPS-data class can be connected to a given PDA and has an identifier and a set of attributes like the latitude and the longitude.
- **Mean of transport**

While moving, the MIC uses a mean of transport. In our model, the mean of transport class has an identifier and some attributes like the color, the minimum speed, the maximum speed.
- **Delegation**

Each stop is in a given delegation. This latter is in a regional government (class) which is in a country (class).
This class has an identifier which is the name of the delegation and others attributes such as surface of the delegation, the population of the delegation and the climate.
- **Points of interest**

We found around each stop several spatial features [36] (a spatial feature type is a real world entity that has a location on the Earth surface (OGC1999)). In our definition, spatial feature types are points of interest. Those latter can be of type Natural or Artificial.
- **Natural**

The class natural is composed by the Classes Sea, Desert, Mountain and Lake.
Each class which is a specialization of the class Natural has an identifier and some specific attributes such as the surface, the length, the depth…

- **Artificial**

The class Artificial is composed by different classes such as Educational-company, Industrial-company, Agricultural-company, Transportation-company, Healthcare-company, Touristy-company, Cul-art-company… Each class has an identifier and others specific properties such as the name, the location, the activity, the type.

**3.2.2. Mobile information collectors' itinerary diagram**

The itinerary describes the tasks of the MIC and the places where those tasks are to be performed. This description defines a static travel planning. We propose the following static itinerary diagram.

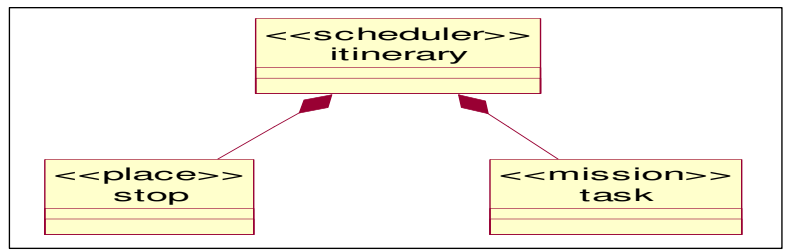

Figure 4. The MIC'static itinerary diagram

According to planned assignment, each MIC navigates, through a defined trajectory (set of stops and moves) during a time interval in order to do some tasks.
While traveling many events may occur and the MIC can not reach a given place, so he can not accomplish some tasks. In order to solve this problem, we propose to define a list of navigation events (a navigation event represents the unexpected events that can be produced during the travel of the MIC) and a list of equivalent places (stops). Then the travel of the MIC becomes dynamic and flexible, since it can be adapted to the environment and the network changes.
To model navigation events and equivalent places, we define a diagram, called Dynamic Itinerary Diagram.

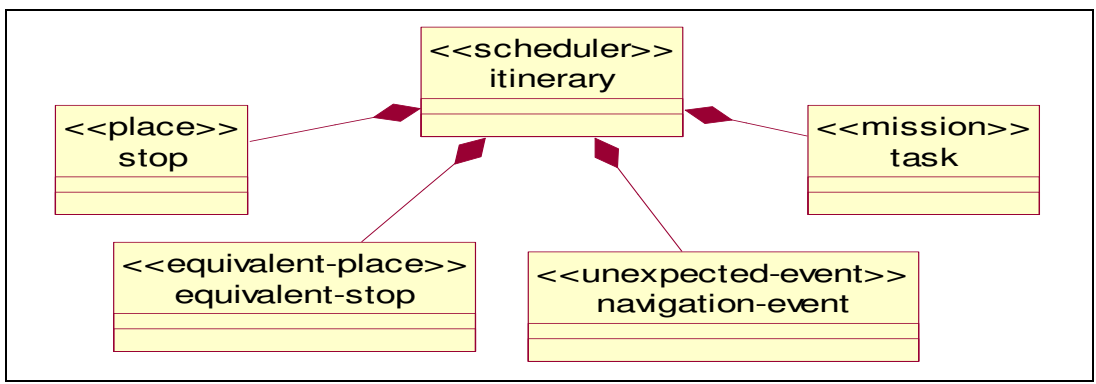

Figure 1. The MIC's dynamic itinerary diagram

Let's mention that the set of navigation events can be updated by the MIC if he met problems during his trajectory.

### 3.2.3. Mobile professionals' trajectory description

The requirements for the trajectory modeling are to take into account the characterization of trajectories and their components and the different types of constraints (semantic, topologic...) in order to fix a conceptual view of the concept of trajectory. The conceptual model is seen as a direct support and an explicit representation of trajectories' components (stops, moves).
For this reason, we propose a navigation diagram in which we are interested in the destination stops and equivalent stops if there are navigation events. The set of tasks in a given stop are not considered in this diagram.
We propose that the delegation model stops and the transition between stops model the moves and problems which trigger the transition between stops model the navigation events. Those latter drives the MIC to equivalent stops.

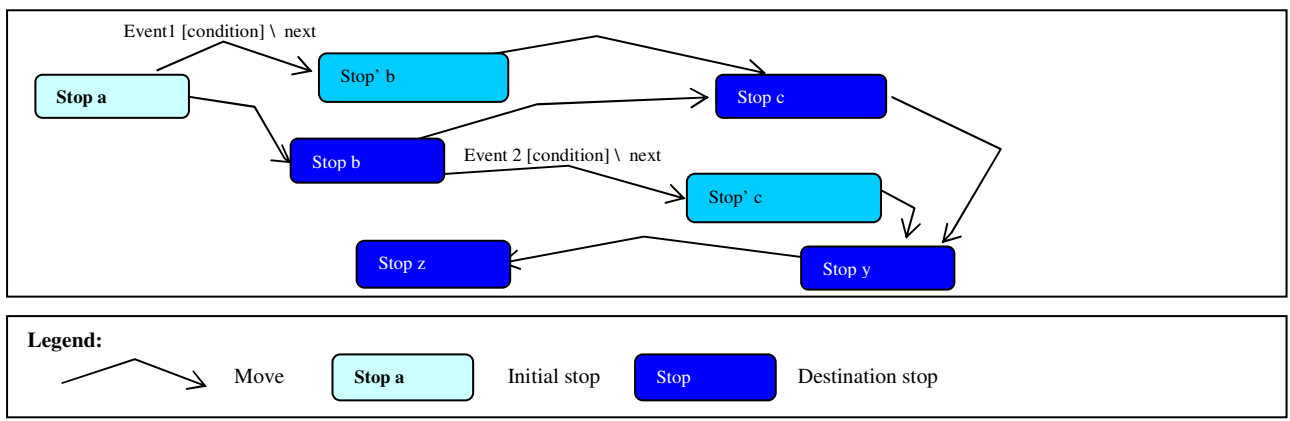

Figure 2. The MIC's navigation diagram

While stopping, the Mobile Professionals have three tasks to do: to observe points of interest, to collect information and to send trajectory data from the same stop or equivalent stop. Those tasks have to be done at each stop belonging to the trajectory of a given MIC.

## 4. TRAJECTORY DATA WAREHOUSE CONCEPTUAL MODELING FOR MICs

In this section we will present Trajectory Data Warehouse modeling for our running example. In fact, we choose as a multidimensional model the snowflake schema. This latter is a variant and a refinement of the star schema, since it keeps the same primitives with normalization of dimension tables to eliminate redundancy. In fact, it represents aggregation hierarchies in the dimensions since, each attribute of a hierarchical level is putted in a dimension table. The snowflake schema [36-37] may improve in some cases the performance because smaller tables are joined, and is easy to maintain and increases flexibility.

The following figure represents the snowflake schema of our application.

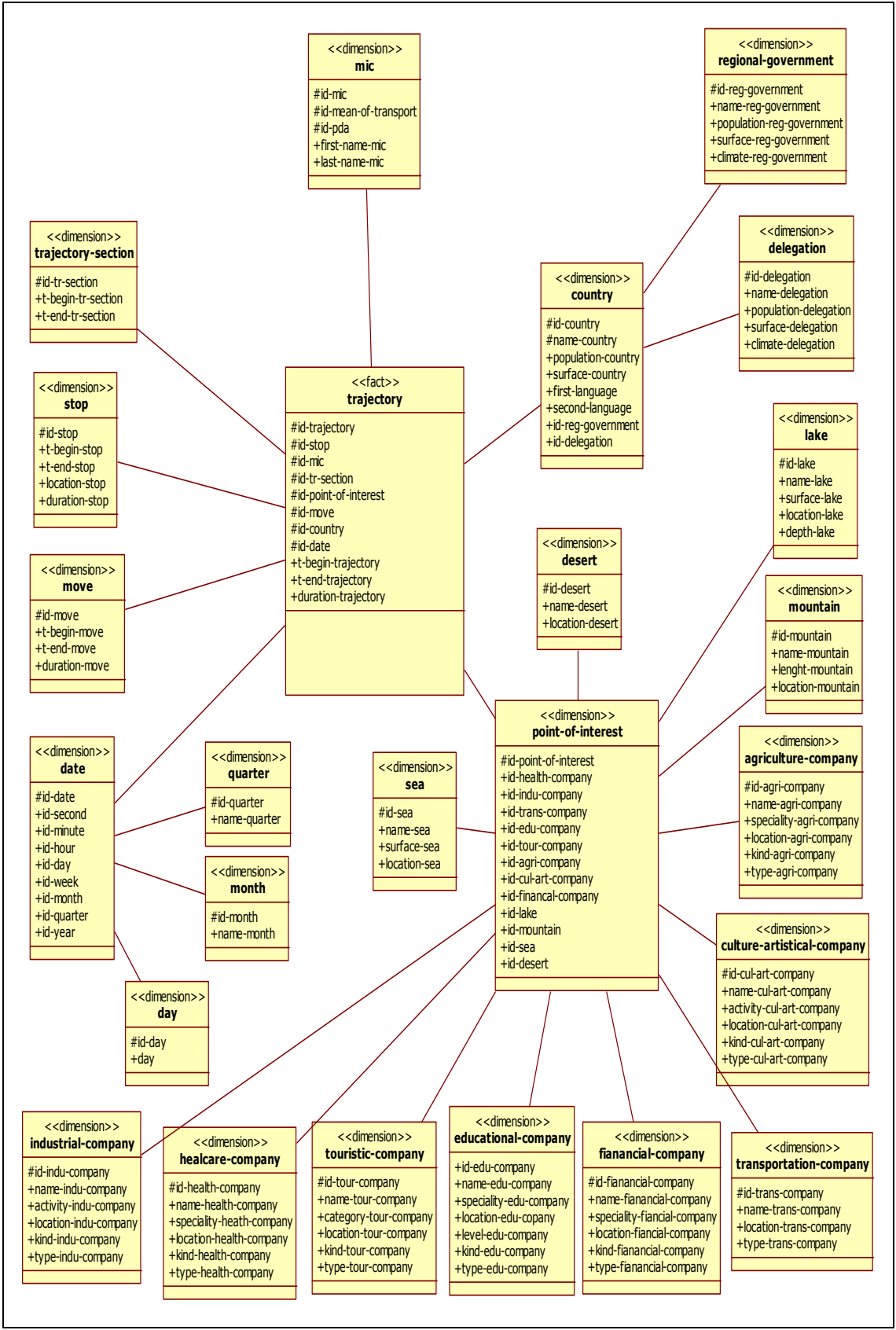

Figure 7. The MIC's trajectory data warehouse snowflake schema

For our application we choose to use the snowflake schema for normalization reasons. In fact, the dimension 'date' can be detailed with others dimension tables like 'month', 'day-of-week' and 'quarter'. The dimension Country can be detailed with two dimension tables like 'delegation' and 'regional-government'. The dimension 'points-of-interest' can be detailed with thirteen dimension tables like 'sea', 'lake', 'mountain' and 'desert' 'educational-company', 'transportation-company', 'industrial-company', 'healthcare-company', 'touristic-company','culture-artistical-company', 'agriculture-company' and 'fianancial-company'.

The snowflake schema in figure 7 is composed by a trajectory fact table describing some measures such as: t-begin-trajectory (to give the time of beginning of a given trajectory), t-end-trajectory (to give the time of end of a given trajectory) and duration-trajectory, and the following dimension classes:

Mic: contains information about mobile information collector like first name, last name.

Date: contains the date specificity like the day, the month, the year, the quarter, the hour.

Move: contains information about the time of begin and of end for a given move.

Tr-section: contains information about the time of begin and of end for a given trajectory section.

Stop: contains information about the time of begin and of end for a given stop and the location.

Country: contains information about a given country like the name, the population.

Delegation: contains information about a given delegation like the name, the population.

Point-of-interest: contains information about different kind of points of interest. In fact, we classify two kinds of points of interests which are artificial and natural. As artificial points of interest we cite: Educational-company,Transportation-company,industrial-company,Healthcare-company,Touristic-company,Cult-art-company,Agriculture-company,Financial-company.  As natural points of interest we cite: Mountain, Sea, Lake, Desert .

The choice of the snowflake schema as a conceptual model for the TrDW of our application does not mean that it is the best model. In fact, the increase of tables occurs the increase of joins. This obviously can affect the response time of queries. Because of some reasons citing before, we opted for the snowflake schema at the conceptual level. At the logical  level  we worked with SQL server 2005 in order to transform the previous snowflake schema into the following schema:

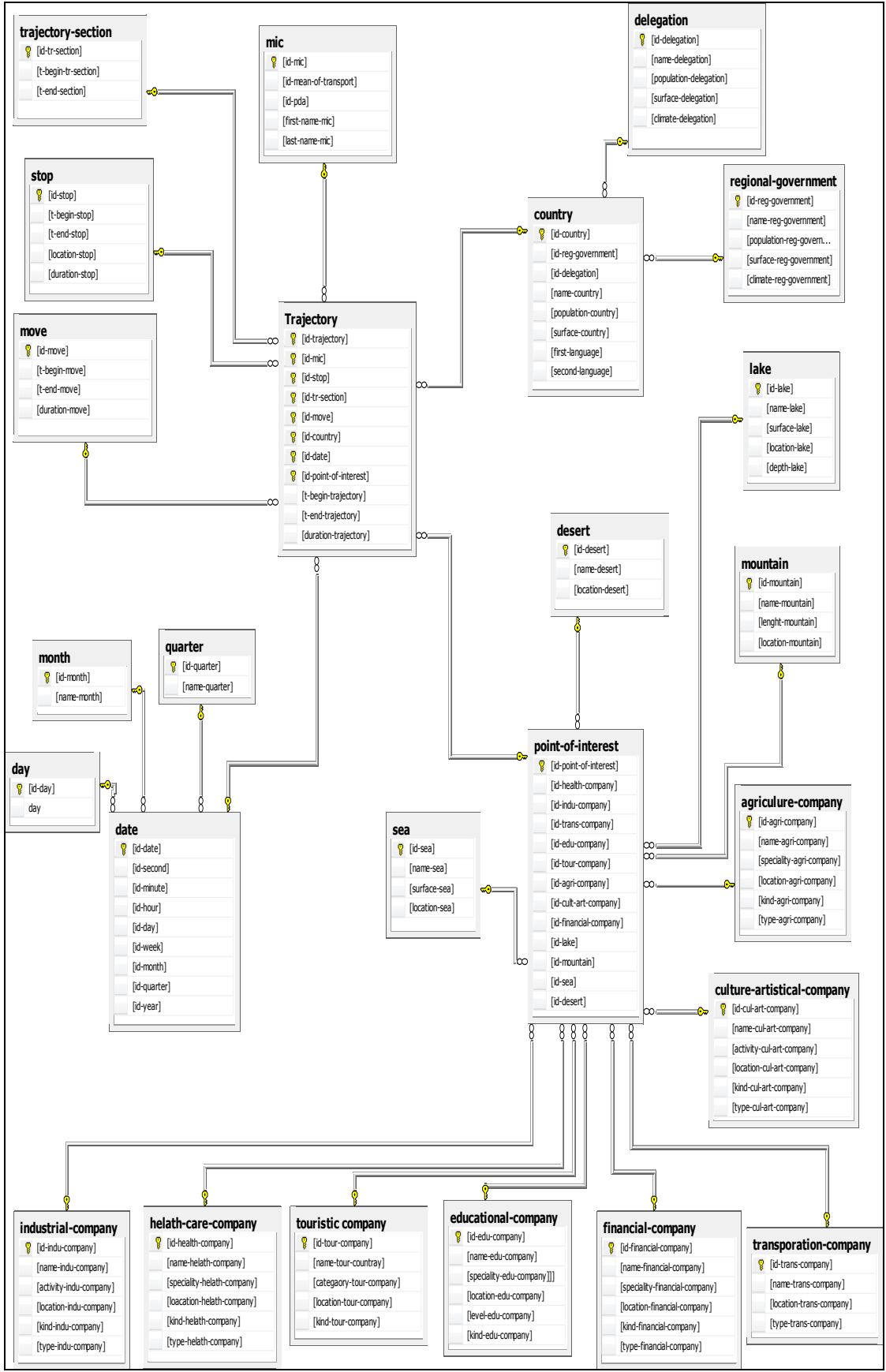

Figure 8. The MIC's trajectory data warehouse

# 5. ANALYSIS

Let's mention that our application is interested in finding new opportunities for new commerce. In fact, in this section, we will define some queries about different existing points of interest and their localization in order to help investors in the choice of the suitable investment.

- Which Tunisian places in the trajectory 34 that contain touristic projects?

Select location-tour-company From  trajectory T, touristic-company TC, point-of-interest P Where T.id-point-of-interest=P.id-point-of-interest and T.id-trajectory=34 and exists (select * from trajectory T' where ( T'.id-trajectory=T.id-trajectory) and T'.id-point-of-interest In (select id-point-of-interest From point-of-interest Where id-tour-company Is Not Null and T'.id-country In (select id-country From country Where name-country="Tunisia"))).

- How many agriculture projects are at sousse?

Select  count(id-agri-company) From  agriculture-company A, point-of-interest P, trajectory T,  country C, delegation D Where T.id-country=C.id-country and C.name-country="tunisia"  and  T.id-point-of-interest=P.id-point-of-interest  and  P.id-agri-company=A.id-agri-company and C.id-delegation=D.id-delegation  and  C.name-delegation="sousse".

- Which places in the trajectory 20 that contain a lake?

Select L.location-lake From lake L Where exists (select * From Trajectory T, Point-of-interest P Where T.id-Trajectory= 20 and T.Point-of-interest= P.Point-of-interest and P.id-lake= L.id-lake).

- Which trajectories that contain a sea and touristic projects?

Select id-trajectoryFrom trajectory T, point-of-interest P, sea S, touristic-company TC Where T.id-point-of-interest=P.id-point-of-interest and P.id-sea=S.id-sea and P.id-tour-company= TC.id-tour-company Ordered by T.id-trajectory.

- Which hotels at Hammamet are 5 stars hotel?

Select name-tour-company,id-tour-companyFrom touristic-company Where category-tour-company="hotel" and type.category-tour-company="5stars" and location-tour-company= Hammamet.

- Which trajectories that contain more than 10 touristic projects?

Select  distinct (T.id-trajectory From trajectory T Where 10< ( select count (*) From Trajectory T' Where T.id-trajectory= T'.id trajectory and T'.id-point-of-interest In ( select P.id-point-of-interest From Point-of –interest P and P.id-tour-company Is not Null)).

At the end of this section we will give an example of an MDX query generated by the analysis service which is a component of SQL server business intelligence development studio.

The following figure is the result of the MDX query which want to give for each country and delegation the number of industrial companies and their number.

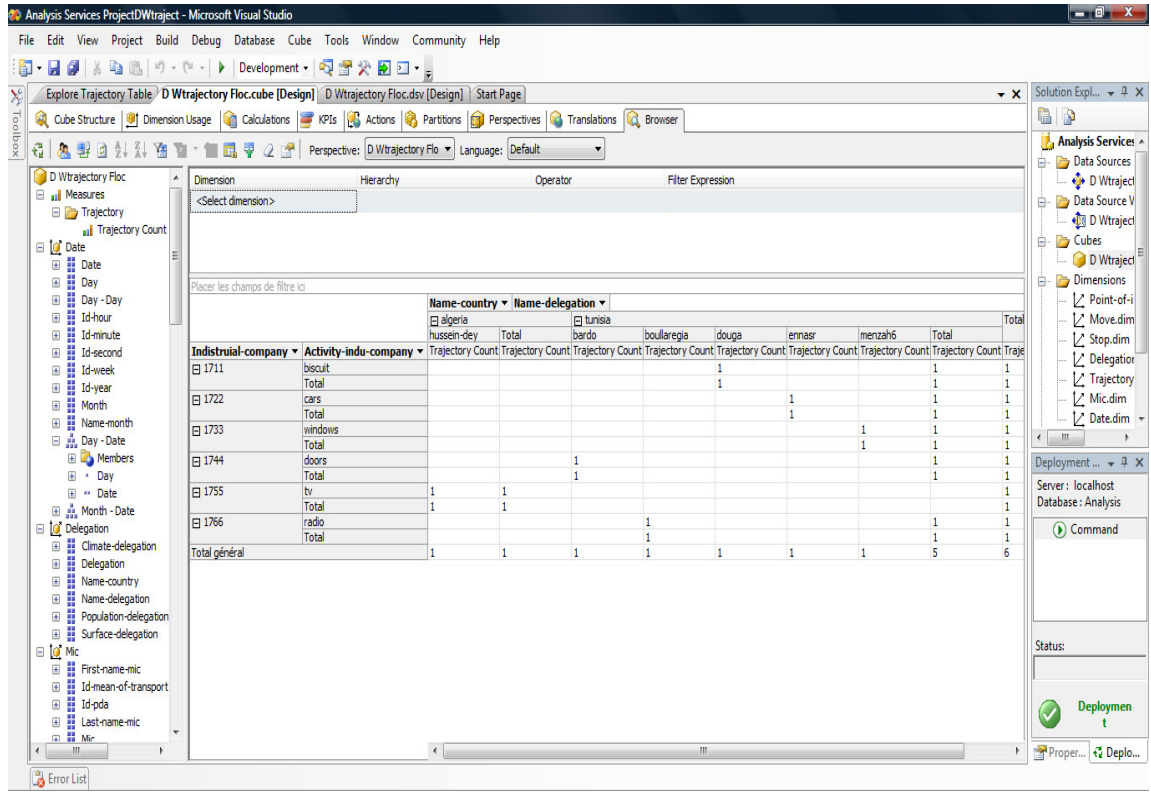

Figure 9. The MDX query

## 6. CONCLUSIONS

In the last years, mobile objects are gaining great attention as a new concept for developing and implementing mobile and distributed applications. However, very little work has taken place in modeling such systems and up to now, none of the existing works were interested in the conceptual modeling of the mobile professionals' trajectory.

In this work, we are presenting the mobile object as a mobile information collector. This latter is involved by a company in collecting and sending data about different points of interest via a PDA. Such data is called trajectory data and it has to be stored in a trajectory data warehouse in order to analyze them at a first step and make strategic decisions about implanting new commercial activities and founding new opportunities at a second step.

As future work, improvements are planed for trajectory data warehouse modeling and using of data mining tools to analyze more deeply the trajectory data in order to perform the decision making and developing Trajectory OLAP that takes into account the continuity of changes of mobile objects in time and space.

## ACKNOWLEDGEMENTS

we would like to thank the High Institute of Management of Tunis especially my supervisor Dr Akaichi Jalel and my familly especially my father Oueslati Abdellatif and my mother Ben Moussa Aicha !